\documentclass[aps,floats]{revtex4}
\usepackage{amsmath}
\usepackage{graphicx,epsfig}

\begin{document}

\def\oppropto{\mathop{\propto}}
\def\opsimeq{\mathop{\simeq}}
\def\opoverderline{\mathop{\overline}}
\def\operarrow{\mathop{\longrightarrow}}
\def\opsim{\mathop{\sim}}

\def\fig#1#2{\includegraphics[height=#1]{#2}}
\def\figx#1#2{\includegraphics[width=#1]{#2}}

\title{ Low-temperature properties of some disordered systems  \\
from the statistical properties of nearly degenerate two-level excitations}

\author{C\'ecile Monthus}
\affiliation{Service de Physique Th\'eorique,
Unit\'e de recherche associ\'ee au CNRS, \\
DSM/CEA Saclay, 91191 Gif-sur-Yvette, France}

\author{Pierre Le Doussal}
\affiliation{CNRS-Laboratoire de Physique Th\'eorique de l'Ecole
Normale Sup\'erieure, \\ 24 rue Lhomond, F-75231
Paris, France}

\begin{abstract}

The thermal fluctuations that exist at very low temperature
in disordered systems are often attributed to
the existence of some two-level excitations.
In this paper, we revisit this question via the explicit studies
of the following 1D models
(i) a particle in 1D random potentials
(ii) the random field Ising chain with continuous disorder distribution.
In both cases, we define precisely the `two-level' excitations
and their statistical properties, and we
show that their contributions to various observables
are in full agreement at low temperature with the
the rigorous results obtained independently.
 The statistical properties of these two-level excitations
moreover yield simple identities at order $T$ in temperature
for some generating functions of thermal cumulants.
For the random-field Ising chain,
in the regime where the Imry-Ma length is large,
we obtain that the specific heat is dominated by small non-universal
excitations, that depend on the details of the disorder distribution,
whereas the magnetic susceptibility and the Edwards-Anderson order parameter
are dominated by universal large excitations,
whose statistical properties only depend on the variance
of the initial disorder via the Imry-Ma length.

\end{abstract}

\maketitle

\widetext


\section{Introduction}

\subsection{Disordered systems at very low temperature }

One of the most important feature of disordered systems is
that they may present states that have an energy
very close to the ground state energy but which are very different
from the ground state in configuration space.
For the spin-glasses, the debate between the droplet
and replica theories concerns the
probabilities and the properties of these states. In the droplet theory
\cite{droplet}, the low-temperature physics is described
in terms of rare regions with nearly degenerate excitations
which appear with a probability that decays with a power-law of their size.
In the replica theory \cite{replica}, the replica symmetry breaking
is interpreted as the presence of many pure states in the thermodynamic limit,
i.e. the nearly degenerate ground states appear with a finite probability
for arbitrary large size.

More generally, the statistical properties of the nearly degenerate 
excitations (their numbers, their sizes, their geometric properties, 
the barriers separating them, etc...)
are interesting in any disordered system, since they govern
all properties at very low temperature.
In particular, a linear behavior in temperature of the specific heat
\begin{eqnarray}
C(T) = b T +O(T^2)
\label{clinear}
\end{eqnarray}
seems rather generic for a large class of disordered models,
ranging from spin-glasses where this behavior is measured experimentally
\cite{binderyoung} and numerically \cite{jerome04}, 
to one-dimensional classical 
spin models where this behavior can be
exactly computed via the Dyson-Schmidt method \cite{livreluck}.
This linear behavior of the specific heat has been also
obtained recently  for classical disordered
elastic systems via the replica variational method \cite{schehr}.
In this respect, the infinite-ranged spin glass models
are anomalous since the first term of the specific heat is only
of order $T^2$ \cite{infinite,houches}.
The leading term of the specific heat at low temperature
has also been studied recently in various quantum disordered models,
and is still under debate between linear $C \sim T$ \cite{quantum1}, quadratic 
$C \sim T^2$ \cite{quantum2} and cubic $C \sim T^3$ \cite{quantum3}.

The linear behavior of the specific heat (\ref{clinear}) is of course
reminiscent of the same behavior in ordinary glasses,
where the usual description involves a phenomenological
theory of two-level systems \cite{houches,philmag,twolevel}. 
Whereas the identification
of these two-level systems in a microscopic theory has remained
problematic for structural glasses (see \cite{heuer} for recent discussions), this problem seems simpler in
disordered models, where concrete proposals have been made.
In the context of spin-glasses for instance
\cite{binderyoung}, it has been proposed a long time ago
that the two-level excitations
are clusters of spins that can be flipped
with respect to the ground state, and that the associated joint probability
$N(\epsilon,v)$ of the energy $\epsilon$ and barrier $v$
determines various static and dynamical properties
at very low temperature \cite{twoleveldasgupta}. However, the computation
of this probability distribution $N(\epsilon,v)$ is difficult
except for very small clusters containing only a few spins \cite{twoleveldasgupta}.
This is why the statistics of large excitations relies
on some scaling assumptions in the droplet theory \cite{droplet}.
In recent years, there has been a lot of numerical efforts
to characterize the distribution and the topology
of these large low-energy excitations \cite{numerical}.

There exists a special class of disordered models for which
exact remarkable identities for thermal fluctuations
 have been obtained \cite{identities88}.
The Hamiltonian of these models have a deterministic part which
consists in quadratic interactions and a random part whose statistics
is translation invariant.
The simplest of these systems
is the so called
toy model \cite{villain83} defined by the one-dimensional Hamiltonian
\begin{eqnarray}
H_{toy}=\frac{g}{2} x^2 +V(x)
\label{deftoy}
\end{eqnarray}
where the random potential $V(x)$ is a Brownian motion presenting the
correlations
\begin{eqnarray}
\overline{ \left(V(x)-V(y) \right)^2 }= 2 \sigma \vert x-y \vert
\label{corre}
\end{eqnarray}
The statistical tilt symmetry
leads to the following result
for the generating function of thermal cumulants
averaged over the disorder \cite{identities88}
\begin{eqnarray}
 \overline { \ln < e^{- \lambda x } > } && = T \frac{\lambda^2}{2  g}
 \label{gentoy}
\end{eqnarray}
This identity is particularly interesting at very low temperature,
since it predicts a linear behavior in $T$ of the second cumulant.
In a previous work \cite{us_toy}, we have explicitly shown that
the rare configurations with two nearly degenerate minima $\Delta E \sim T$
 actually give the full exact second cumulant of (\ref{gentoy}).
The statistical tilt symmetry is also present in the models
of directed polymers in random media, and we refer the reader
 to the references \cite{huse,hwa}
for a detailed discussion of the identities on thermal fluctuations and their interpretation in terms of nearly degenerate paths.

\subsection{ Goal and results }

The above presentation on the properties of disordered systems
at very low temperature, although very incomplete,
 already shows that generically, various observables exhibit
variations at order $T$ in temperature, that can be related to the statistical properties of nearly degenerate excitations.
The aim of this paper is to revisit this question in some
one-dimensional models to identify precisely the `two-level' excitations
and their statistical properties, in order to compare their contribution
to various observables at order $T$ with
the rigorous results that can be obtained independently.

\subsubsection{ Results for one particle in a random potential}

We will first consider the equilibrium of a particle
in random potentials of the following form
\begin{eqnarray}
H(x)=H_0(x)+V(x)
\label{1Dmodels}
\end{eqnarray}
where $H_0$ is the non-random part of the Hamiltonian, and
where the random potential $V(x)$ is a Brownian potential (\ref{corre}).
As explained above, the toy model
$H_0=\frac{\mu}{2} x^2$ is very special since it satisfies
exact identities (\ref{gentoy}) as a consequence of the statistical tilt symmetry. Here we show that for a large class of deterministic part $H_0$,
the following temperature expansion holds for the generating function
of thermal cumulants
\begin{eqnarray}
 \overline { \ln < e^{- \lambda x } > } && = - \lambda \overline {x_{min}}
+ T \frac{\lambda^2}{2} \int_0^{+\infty} y^2 D(y)
+O(T^2)
 \label{identitylowT}
\end{eqnarray}
where $\overline {x_{min}}$ is the disorder average of the position $x_{min}$
where $H(x)$ is minimum, and where
$D(y)$ is the probability to have two degenerate minima in the system
separated by a distance $y$.
The thermal fluctuations of these models
are thus directly related to the presence of metastable states
in rare disordered samples. In particular, the susceptibility
\begin{eqnarray}
\chi \equiv \frac{1}{T} \left( <x^2>-<x>^2 \right)
\label{suscepti}
\end{eqnarray}
has for average a finite value at zero-temperature
\begin{eqnarray}
\overline{ \chi } = \int_0^{+\infty} y^2 D(y) +O(T)
\end{eqnarray}
but only the samples with two nearly degenerate minima contribute
to this zero-temperature value. The typical samples with only
one minimum have a susceptibility that vanishes at zero temperature.
This type of behavior was found numerically for the random directed polymer
model \cite{mezard}.
We will show that the relation (\ref{identitylowT}) is in full agreement
with the exact results corresponding to the two soluble cases \\
(i) the pure Brownian potential $H_0=0$ on a finite interval $x \in [0,L]$; \\
(ii) the biased Brownian potential $H_0=fx$ on the semi-infinite
 line $x \in [0,+\infty[$.

\subsubsection{ Results for the random field Ising chain }

We will then consider the one-dimensional
random field Ising model, in the regime where the Imry-Ma length
$L_{IM}$ representing the typical domain size at zero temperature is large.
In contrast with models (\ref{1Dmodels}) with only one degree of freedom,
where the low-temperature properties are governed by the rare {\it samples}
having two nearly degenerate ground states, the systems with many degrees of freedom are well described by disorder averaged values in the thermodynamic limit,
since the average over the disorder is like a spatial average.
As a consequence, the low temperature properties are governed by
the rare {\it regions } that presents
nearly degenerate excitations.
We describe in detail the rare nearly degenerate
excitations and evaluate their contribution to the low temperature behavior of various observables.
The density $ \rho(E=0,l) $ of excitations of length $l$
contains two contributions, involving either a single domain wall,
that has two nearly degenerate optimal positions, or
a pair of neighboring domain walls, than can appear or disappear with
almost no energy cost.
Our analysis allows to obtain that the specific heat \cite{philmag}
\begin{eqnarray}
C(T)= T \frac{\pi^2}{6} \int dl \rho(E=0,l) +O(T^2)
\end{eqnarray}
is dominated by small non-universal excitations of length $l \sim 1$,
that depend on the details of the disorder distribution,
whereas the Edwards-Anderson order parameter
\begin{eqnarray}
 q_{EA}= \overline{ <S_i>^2 } = 1- 2 T
 \int_0^{+\infty} dl \ l \rho(E=0,l) +O(T^2)
\end{eqnarray}
and the magnetic susceptibility
\begin{eqnarray}
\overline{\chi} \equiv \frac{N}{T} \left( \overline{ <m^2>-<m>^2 } \right)
= 2
\int_0^{+\infty} dl \ l^2 \rho(E=0,l) +O(T)
 \end{eqnarray}
are dominated by large excitations whose length $l$ is of order of the Imry-Ma length $L_{IM}$, and whose properties are universal with respect
to the initial disorder distribution, since they only depend upon its variance.
 We will then compare with the exact results available
from the Dyson-Schmidt method \cite{theoluck}.

\subsubsection{ Organization of the paper }

The paper is organized as follows.
The Section \ref{secparticle} is devoted to the equilibrium at low
temperature in the one-dimensional potentials (\ref{1Dmodels}).
The Section \ref{secrfim} concerns the analysis of two-level excitations
in the random field Ising chain, and their influence on various observables. The Appendix \ref{purepath} contains some exact results on thermal cumulants
in the Brownian potential, that are useful to show the exactness
of the two-level description presented in Section \ref{secparticle}.

\section{ Equilibrium of a particle in one-dimensional random potentials}

\label{secparticle}

\subsection{Effective model at order $T$ in temperature}

\label{2delta}

In typical samples with Hamiltonian (\ref{1Dmodels}),
there is only one minimum.
In \cite{us_golosov}, it was moreover shown that the typical extension
of the Boltzmann distribution around a minimum for a pure Brownian potential
is given by the thermal length $l_T= \frac{T^2}{\sigma}$,
that comes from the Boltzmann factor $e^{- \beta V(x)}$
with the typical behavior $V(x) \sim \sqrt{ \sigma x}$.
For the other models (\ref{1Dmodels}) with regular deterministic part, the leading behavior at small
distance is still given by the Brownian potential $V(x)$, and thus
the same argument can be applied. As a consequence, for all potentials (\ref{1Dmodels}),
the typical fluctuations around a minimum are of order $\Delta x \sim T^2$
at low temperature, and thus do not contribute at order $T$.

\begin{figure}
\centerline{\includegraphics[height=6cm]{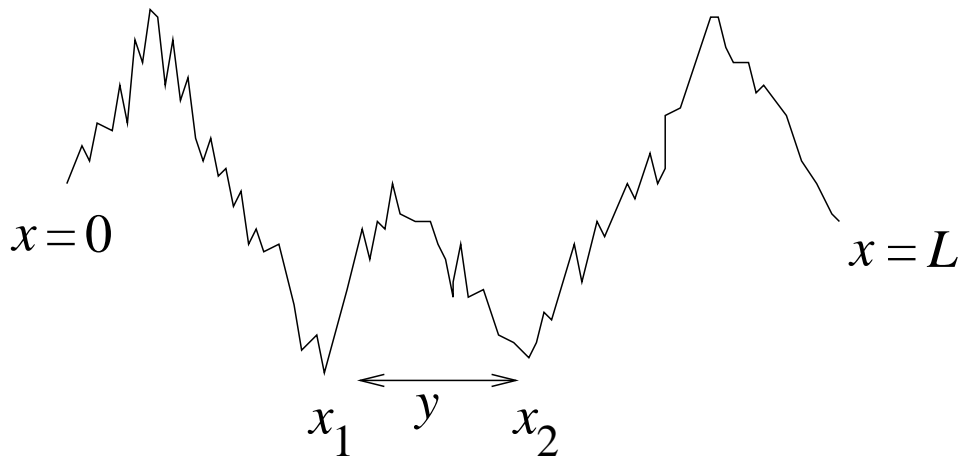}} 
\caption{ Example of a random potential presenting two nearly degenerate minima   situated at the positions $x_1$ and $x_2$. The low temperature equilibrium properties are
completely determined by the probability $D(y)$ over the samples that there
exist two nearly degenerate minima separated by a distance $y$.} 
\label{rareb}
\end{figure}

We now consider the configurations presenting
 two almost degenerate minima $H(x_1) \simeq H(x_2)$
situated at $x_1$ and $x_2$ : see Figure \ref{rareb}.
To compute their contribution at order $T$, we may neglect the spatial extension
of order $l_T \sim T^2$ around each minimum, and replace the thermal measure by two delta peaks
\begin{eqnarray}
e^{- \beta h_{eff} } = p \delta(x-x_1) +(1-p) \delta(x-x_2)
\label{zeff}
\end{eqnarray}
The respective Boltzmann weights $p$ and $1-p$ are given by
\begin{eqnarray}
 p = \frac{ e^{- \beta H(x_1)} } { e^{- \beta H(x_1)} + e^{- \beta H(x_2)} }
= \frac{ 1 } { 1 + e^{ - \beta E} }
\end{eqnarray}
where $E=U(x_2)-U(x_1)$ is the small energy difference between the two minima.
Since we are interested in a window of size $ \vert E \vert < T$, as $T \to 0$,
the probability to have two almost degenerate minima situated at $x_1$ and $x_2$
with an energy difference $E$ can be considered to be
 uniform in the energy leading to the approximation
$\sim P(E=0,x_1,x_2) dE$.
Here we choose the convention that $y=x_2-x_1$ is of arbitrary sign.
Then to avoid double counting, we define $x_1$ as the true minimum,
so that $E \geq 0$.
As in the study of the rare events in \cite{us_sinai,us_rfim},
the average over $E$ of the quantities of interest can
be performed by integrating over $E \in [0,+\infty[$.
More precisely, we introduce the symbol $<<..>>$ to denote
the following average over configurations with two degenerate minima
\begin{eqnarray}
<< A(p,x_1,x_2) >> \equiv \int dx_1 \int dx_2 \int_0^{+\infty} dE
\ P(E=0,x_1,x_2)
 \ \ A \left( p= \frac{ 1 } { 1 + e^{ - \beta E} },x_1,x_2 \right)
\label{stat}
\end{eqnarray}

At this point, it is important to stress here that the measure
 $<<.>>$ is not normalizable, in the sense that $<<1>>$ diverges,
but has only a meaning when the observable $A$ vanishes when only one minimum exists, that is for $p \to 1$ or for $y \to 0$.
 As first examples
of the measure defined above, let us compute the lower cumulants of the position.
The variable
$(x-<x>)$ is $(1-p)(x_1-x_2)$ with probability $p$ and $p (x_2-x_1)$ with probability $1-p$, leading to
\begin{eqnarray}
 c_2(T) && \equiv \overline{ < (x-<x>)^2 > }  = << p(1-p) (x_1-x_2)^2 >> +O(T^2) \\
&& = \frac{T}{2} \int dx_1 \int dx_2  P(E=0,x_1,x_2)  (x_1-x_2)^2
\label{two}
\end{eqnarray}
Similarly, the fourth cumulant
\begin{eqnarray}
c_4(T) \equiv \overline{ < (x-<x>)^4     - 3  < (x-<x>)^2 >^2 > }
\end{eqnarray}
can be computed as follows
\begin{eqnarray}
c_4(T)  = << p (1-p) (1 - 6 p + 6 p^2) (x_1-x_2)^4 >>  +O(T^2)  =0 +O(T^2)
\label{four}
\end{eqnarray}
The vanishing result is thus the consequence of a special cancellation
of two terms of order $T$.
These cancellations actually occur at all orders, as
already noted in \cite{huse} for the random directed polymer model.
We now present a proof within the effective model (\ref{zeff}).

\subsection{Generating function of thermal cumulants at order $T$ in temperature}

We now consider the
partition function of the effective hamiltonian (\ref{zeff}) with a source
$\lambda$
\begin{eqnarray}
Z_{eff}(\lambda) = \int dx e^{-\beta h_{eff} - \lambda x} = p e^{-\lambda x_1}
+(1-p) e^{-\lambda x_2}
\end{eqnarray}
The generating function of disorder averages of the thermal cumulants
can be now expanded in temperature as as
\begin{eqnarray}
\overline{  \ln  Z_{eff} (\lambda) } && =   - \lambda  \overline{x_1}
 + << \ln \left(p
+(1-p) e^{-\lambda (x_2-x_1)} \right) >> +O(T^2)
\end{eqnarray}
Since $x_1$ has been defined as the true minimum $x_{min}$
of the Hamiltonian $H(x)$,
and since the variable $y=x_2-x_1$
has zero odd moments, we may rewrite
\begin{eqnarray}
\overline{ \ln  Z_{eff} (\lambda)} && =  - \lambda  \overline{x_{min}}
+\frac{1}{2}
 \left[ << \ln \left(p  +(1-p) e^{\lambda y} \right)
+ \ln \left(p  +(1-p) e^{-\lambda y} \right) >>  \right] \\
&& =  - \lambda  \overline{x_{min}} + \frac{1}{2} \int_{-\infty}^{+\infty} dx_1 \int_{-\infty}^{+\infty} dy P(E=0,x_1,x_1+y)
\int_0^{+\infty} dE  \ln [ \frac{ \left(1  + e^{- E/T} e^{\lambda y} \right)
\left(1  + e^{- E/T} e^{-\lambda y} \right) }{ (1 + e^{- E/T})^2 } \nonumber
\end{eqnarray}
It turns out that the integral over the energy difference variable $E$
yields the very simple result
\begin{eqnarray}
\int_0^{+\infty} dE  \ln [ \frac{ \left(1  + e^{- E/T} e^{\lambda y} \right)
\left(1  + e^{- E/T} e^{-\lambda y} \right) }{ (1 + e^{- E/T})^2 }
 = T \int_0^{1} \frac{dw}{w}
\ln [ \frac{ \left(1  + w^2+2 w \cosh(\lambda y) \right) }{ 1+w^2+2 w }
 = \frac{T}{2} \lambda^2 y^2
\label{integralenergy}
\end{eqnarray}
Using the symmetry in $y \to -y$, the final result reads
\begin{eqnarray}
\overline{ \ln  Z_{eff} (\lambda)} && = - \lambda  \overline{x_{min}}
+ \frac{T}{2} \lambda^2
 \int_{0}^{+\infty} dy y^2 D(y) +O(T^2)
\end{eqnarray}
where
\begin{eqnarray}
D(y) = \int_{-\infty}^{+\infty} dx_1 P(E=0,x_1,x_1+y)
\label{defD}
\end{eqnarray}
represents the probability to have two degenerate minima separated by a distance $y \geq 0$.

The conclusion is thus that the exact identities for the thermal cumulants
(\ref{gentoy}) derived from the statistical `tilt' symmetry \cite{identities88}
are in fact true at order $T$ for all the models (\ref{1Dmodels})
that do not present the statistical tilt symmetry.

\subsection{Disorder averages of the moments at order $T$ in temperature}

Similarly, the even moments of $(x-<x>)$
have the following small-T expansion
\begin{eqnarray}
 \overline {  < (x-<x>)^{2k} > } && = << \left[ p(1-p)^{2k} +
+ (1-p) p^{2k} \right] (x_1-x_2)^{2k}  >>  +O(T^2) \\
&& = \frac{T}{k} \int_0^{+\infty} y^{2k} D(y) +O(T^2)
\label{momentslowT}
\end{eqnarray}
where $D(y)$ is defined as in (\ref{defD}).
So the whole distribution $D(y)$ is important
to characterize the low temperature properties.
The comparison with (\ref{identitylowT}) shows that there are
a lot of cancellations in the disorder averages of the cumulants.
This phenomenon was already emphasized in \cite{huse}
for the random directed polymer.

\subsection{`Chaos' in position at order $T$ in temperature}

Another interesting observables are the `chaos observables'
that contain information about the configurational changes induced by temperature shifts. For instance, among the various chaos observables
 that have been discussed for the
Sinai model \cite{salesjpb}, the following quantity
\begin{eqnarray}
d_{T_1,T_2} = \overline{ (< x >_{T_1} - < x >_{T_2})^2 }
\end{eqnarray}
characterizes the change in the mean thermal position
between two temperatures.
Again, at order $T$ in temperature, this quantity
will be governed by
configurations with two nearly degenerate minima
\begin{eqnarray}
d_{0,T} = << (1-p)^2 (x_2-x_1)^2 >>  +O(T^2) = (2 \ln 2-1) \int_0^{+\infty} y^{2} D(y) +O(T^2)
\label{chaosordret}
\end{eqnarray}
More generally, higher moments have the following temperature expansion
\begin{eqnarray}
\overline{ (< x >_{T} - < x >_{T=0})^{2k} }
= << (1-p)^{2k} (x_2-x_1)^{2k} >>  +O(T^2) = T m_k \int_0^{+\infty} y^{2k} D(y) +O(T^2)
\end{eqnarray}
where
\begin{eqnarray}
 m_k = 2 \int_0^1 dw \frac{w^{2k-1}}{(1+w)^{2k} }
\end{eqnarray}

One can also compute, in the limit where $T_1/T_2=z$ is fixed and $T_1 \to 0$
\begin{eqnarray}
\overline{ (< x >_{T_1} - < x >_{T_2})^{2k} } = T_1 M_k(T_1/T_2)
\int_0^{+\infty} y^{2k} D(y) 
\end{eqnarray}
with
\begin{eqnarray}
M_k(z) = 2 \int_0^1 \frac{dw}{w} \frac{(w - w^z)^{2k}}{(1+w)^{2k} (1+w^z)^{2k} }
\end{eqnarray}
In particular, for a small change $\delta T \ll T$, the change in position is characterized by the moments
\begin{eqnarray}
\overline{ (< x >_{T} - < x >_{T+ \delta T})^{2k} } \sim 2 T  (\frac{\delta T}{T})^{2 k} 
\int_0^1 dw \frac{w^{2k-1}}{(1+w)^{4k} }
(\ln w)^{2 k} \int_0^{+\infty} y^{2k} D(y)  
\end{eqnarray}

\subsection{Discussion}

All quantities at order $T$ in temperature may
 thus be exactly computed with the simple effective thermal measure (\ref{zeff})
where the statistics of the parameters $(x_1,x_2,p)$
is defined by (\ref{stat}).
For the case of the `toy model' (quadratic well plus Brownian potential),
which present the `statistical tilt symmetry'
we have already studied in \cite{us_toy} the configurations
with two nearly degenerate minima and showed that they give the exact result
at order $T$ for the generating function of thermal cumulants.
We will now consider models which do not present the statistical tilt symmetry.

\subsection{Explicit study of the Brownian potential on finite sample}

\label{brownian}

In this section, we consider a pure Brownian potential,
 i.e. $H_0=0$ in (\ref{1Dmodels}), on a finite sample $x \in [0,L]$.
Many properties of this model have already been studied
\cite{broderix-kree,oshanin,flux,yor,us_golosov}. We first derive
the exact thermal cumulants of the position, and we then compare with
the contribution of configurations presenting two nearly degenerate minima.

\subsubsection{Exact temperature series expansion for the thermal cumulants of the position}

The generating function of the disorder averages of thermal cumulants is  given by
\begin{eqnarray}
 \overline { \ln < e^{- \lambda x } > } &&
 =  \overline {\ln Z_L(\lambda)} - \overline {\ln Z_L(0)}
 \label{defgene}
\end{eqnarray}
where $Z_L(\lambda)$ represents the partition function with source $\lambda$ \begin{eqnarray}
 Z_L(\lambda) = \int_0^{L} dx e^{- \lambda x} e^{-\beta V(x)}
\label{zlambda}
\end{eqnarray}
The full probability distribution of this partition
function is exactly known \cite{flux,yor}.
For our present purposes, it is simpler to work with the following
formula in Laplace transform with respect to the length $L$ \cite{yor}
\begin{eqnarray}
\omega \int_0^{+\infty} dL e^{- \omega L} \overline{ \ln Z_L(\lambda)  }
&& = \ln  \left( \frac{1}{\beta^2 \sigma }\right)+\psi(1)
+ \frac{2}{\nu^2} \left( \sqrt{\nu^2+a^2}-a \right) \nonumber \\
&& - \left[ \psi\left(1+ \frac{\sqrt{ \nu^2 + a^2}}{2} + \frac{a}{2}  \right)
+ \psi \left(1+ \frac{\sqrt{ \nu^2 + a^2}}{2} - \frac{a}{2}\right) \right]
\label{reslaplace}
\end{eqnarray}
where $\psi(x)= \frac{\Gamma'(x)}{\Gamma(x)}$ with the following notations
\begin{eqnarray}
a && = \frac{\lambda}{\sigma \beta^2} \\
\nu && = \frac{2}{\beta}
 \sqrt{\frac{\omega}{\sigma}}
\label{nota}
\end{eqnarray}
For completeness, a derivation of (\ref{reslaplace}) via path-integrals
is given in Appendix \ref{purepath}.
To obtain the thermal cumulants, we need to perform a series expansion in
$\lambda$, i.e. in $a= \frac{\lambda}{\sigma \beta^2}$ in formula
(\ref{reslaplace}). Since the only term odd in $\lambda$
 on the right hand-side is simply $(- a \frac{2}{\nu^2})$, we get
\begin{eqnarray}
\overline{ <x> } = \frac{L}{2}
\end{eqnarray}
as expected by symmetry, and
all other odd cumulants identically vanish, as a consequence
of the symmetry $x \to L-x$
 of the random potential $V(x)$.
We now consider the even cumulants. The cumulant of order two is given by
\begin{eqnarray}
\omega \int_0^{+\infty} dL e^{- \omega L} \overline{ (<x^2>-<x>^2) }
&& = \frac{1}{(\sigma \beta^2)^2}  \left[ \frac{2}{\nu^3}
-  \frac{1}{\nu} \psi' \left( 1+ \frac{\nu}{2} \right)
- \frac{1}{2} \psi'' \left( 1+ \frac{\nu}{2} \right) \right] \\
\end{eqnarray}
We may now expand in $\nu$ (\ref{nota}) to obtain the small temperature
 expansion, which also corresponds to the large $L$ expansion
\begin{eqnarray}
\int_0^{+\infty} dL e^{- \omega L} \overline{ (<x^2>-<x>^2) }
&& = \frac{T^4}{ \omega \sigma^2 }  \left[ \frac{2}{\nu^3}
-  \frac{\pi^2}{6\nu} - \psi''(1) +O(\nu) \right] \\
&& = \frac{T}{ 4 {\sqrt \sigma} \omega^{5/2} }
- \frac{\pi^2 T^3 }{12 \sigma^{3/2} \omega^{3/2}}
- \frac{\psi''(1) T^4 }{ \sigma^2 \omega}  +O(\frac{T^5}{\sigma^{5/2} \omega^{1/2}})
\end{eqnarray}
After Laplace inversion, the final result reads
\begin{eqnarray}
c_2(T) \equiv \overline{ (<x^2>-<x>^2) }
 = \frac{T}{3 {\sqrt{ \pi  \sigma} }  }
 L^{3/2}  - \frac{\pi^{3/2} T^3 L^{1/2} }{ 6 \sigma^{3/2} }
- \frac{\psi''(1) T^4 }{ \sigma^2 }  +O(\frac{T^5}{\sigma^{5/2} L^{1/2}})
\end{eqnarray}

Similarly, the cumulant of order $4$ has the following low-temperature expansion
\begin{eqnarray}
 c_4(T) && \equiv  \overline{ (<x^4>-4 <x^3><x> -3 <x^2>^2+12 <x^2><x>^2-6 <x>^4) } \\
&&  = - \frac{ T^3 L^{5/2} }{ 10 {\sqrt \pi} \sigma^{3/2} }
 + \frac{ \pi^{3/2} T^5 L^{3/2} } {12 \sigma^{5/2} }
- \frac{ \pi^{7/2} T^7 L^{1/2} }{8 \sigma^{7/2} }
- \psi''''(1) \frac{ T^8  }{\sigma^4} +O( \frac{T^9}{\sigma^{9/2} L^{1/2}} )
\end{eqnarray}

More generally, the leading contribution to the cumulant of order $(2 k)$
reads :
\begin{eqnarray}
\overline{c_{2k}} (T) && =  \frac{(-1)^{k+1} \Gamma(k-\frac{1}{2})}{
\pi (2k+1) \sigma^{k-1/2}  }T^{2k-1} L^{k+1/2} +...
\end{eqnarray}

The final result is thus that at low temperature
\begin{eqnarray}
 \overline{ \ln <e^{- \lambda x} >  }
= - \lambda \frac{L}{2} + \frac{T \lambda^2}{6 {\sqrt{ \pi  \sigma} }  }
 L^{3/2} +O(T^3)
\label{resbrownian}
\end{eqnarray}
Note that here the next correction is of order $T^3$.
We will now show that this result at order $T$ exactly comes from
the configurations with two degenerate minima.

\subsubsection{Study of configurations with two degenerate minima}

It is straightforward to obtain that
the probability to have two degenerate minima situated at $x_1<x_2$
for a Brownian potential on $[0,L]$ reads
\begin{eqnarray}
P(E=0,x_1,x_2)
= \frac{1}{2 \pi^{3/2} \sqrt{\sigma} \sqrt{x_1} (x_2-x_1)^{3/2} \sqrt{L-x_2}}
\end{eqnarray}

The total probability that it happens with a distance
 $y$ between the two minima
reads
\begin{eqnarray}
D(y) = \theta(L-y) \int_0^{L-y} dx_1 P(E=0,x_1,x_1+y)
=  \frac{ \theta(L-y) }{ \sqrt{ 4  \pi \sigma } y^{3/2} }
\label{dybrownian}
\end{eqnarray}
with the following moments
\begin{eqnarray}
\int_0^L dy y^n D(y)
=  \frac{ L^{n-\frac{1}{2}} }{2 \sqrt{  \pi \sigma  }
 \left(n-\frac{1}{2} \right) }
\label{momentsD}
\end{eqnarray}
This means that configurations with two degenerate minima appear with
a probability of order $1/{\sqrt {\sigma L}} $
and that the distance between
the minima is of order $L$.
In particular, the second moment
\begin{eqnarray}
\int_0^L dy y^2 D(y)
=  \frac{ L^{\frac{3}{2}} }{3 \sqrt{  \pi  \sigma }  }
\end{eqnarray}
exactly corresponds to the result (\ref{resbrownian}) via (\ref{identitylowT}).

The relation (\ref{momentslowT}) for the moments yields
\begin{eqnarray}
 \overline {  < (x-<x>)^{2k} > } = \frac{T}{k} \int_0^{+\infty} y^{2k} D(y)
 =  \frac{T}{k}  \frac{ L^{(4k-1)/2} }{ \sqrt{  \pi \sigma  }
 \left( (4k-1) \right) }
\end{eqnarray}
whereas the relation (\ref{chaosordret}) reads
\begin{eqnarray}
\overline{ (< x >_{T} - < x >_{T=0})^2 } =
T L^{\frac{3}{2}}  \frac{  (2 \ln 2-1) }{3 \sqrt{  \pi  \sigma }  }  +O(T^2)
\end{eqnarray}

\subsection{Explicit results for the biased Brownian potential on the semi-infinite line}

\label{drift}

We now consider (\ref{1Dmodels}) with $H_0=fx$ with $f>0$
on the semi-infinite line $x \in [0,+\infty[$ \cite{opper}.
Again, we compare exact results with the contribution of
configurations with two nearly degenerate minima.

\subsubsection{ Exact results for the thermal cumulants}

The distribution of
\begin{eqnarray}
 Z_f(\lambda) = \int_0^{+\infty} dx e^{- \lambda x} e^{-\beta (fx+V(x))}
= Z_0(\lambda+ \beta f)
\end{eqnarray}
is exactly known \cite{annphys,opper,flux,yor}, and in particular we have \cite{opper,yor}
\begin{eqnarray}
\overline{ \ln Z_f (\lambda) } =  \ln \left(\frac{T^2}{\sigma} \right)
+ \frac{ \sigma }{f T + \lambda T^2 } - \psi \left(1+ \frac{f T }{\sigma} + \frac{\lambda T^2 }{ \sigma} \right)
\end{eqnarray}
where $\psi(x)=\Gamma'(x)/\Gamma(x)$.

The expansion in $\lambda$ yields the cumulants
\begin{eqnarray}
c_1(T) && \equiv \overline{ <x>  }
   = \frac{\sigma }{  f^2 } +  \frac{T^2 }{ \sigma} \psi'( 1+ \frac{f T }{ \sigma}  ) = \frac{\sigma }{  f^2 }+ \frac{\pi^2 T^2}{6 \sigma}+O(T^3)
\\
c_2(T) && \equiv \overline{ <x^2> -<x>^2 }
=   \frac{2 T \sigma }{  f^3 }- \frac{T^4 }{ \sigma^2}
\psi''( 1+ \frac{f T }{ \sigma}  ) = \frac{2 T \sigma }{  f^3 }- \frac{T^4 }{ \sigma^2} \psi''( 1 ) +O(T^5) \\
c_3(T) && \equiv\overline{ (<x^3>-3 <x^2> <x>+2 <x>^3) }
= \frac{6 \sigma T^2}{f^4} + \frac{\pi^2 T^6}{15 s^3}+O(T^7)
\label{cumbiased}
\end{eqnarray}
More generally, the leading order at small temperature of the cumulant
of order $k$ reads
\begin{eqnarray}
c_k(T) = \frac{ k! \sigma T^{k-1}}{f^{k+1}} + ...
\end{eqnarray}
Finally, at order $T$, the exact series expansion of the generating function
of the cumulants reads
\begin{eqnarray}
 \overline{ \ln <e^{- \lambda x} >  }
= - \lambda \frac{\sigma }{  f^2 }
+ \frac{ T \sigma \lambda^2 }{  f^3 }
  +O(T^2)
\label{resbiased}
\end{eqnarray}
We will now compare with the analysis of configurations
with two nearly degenerate minima.

\subsubsection{Configurations with two degenerate minima}

The distribution of the minimum $x_0$ for the biased Brownian
potential on $[0,+\infty[$ reads
\begin{eqnarray}
P_{\min}(x_0)= \frac{f}{\sqrt{4 \pi \sigma} }
\int_{x_0}^{+\infty} \frac{dx}{x^{3/2}} e^{- \frac{f^2 x}{4 \sigma}}
\end{eqnarray}
In particular, the first moment reads
\begin{eqnarray}
\int_0^{+\infty} dx_0 x_0 P_{\min}(x_0)= \frac{ \sigma}{f^2 }
\end{eqnarray}
in agreement with $c_1(T)$ (\ref{cumbiased}) at $T=0$.

The probability to have two degenerate minima situated at $x_1<x_2$
on $[0,L]$ reads
\begin{eqnarray}
P(E=0,x_1,x_2) = P_{min}(x_1) d_f(x_2-x_1)
\end{eqnarray}
where
the probability $d_f(x)$
of return to the minimum after a distance $x$ reads
\begin{eqnarray}
d_f(x)=  \frac{ 1 }{\sqrt{ 4 \pi \sigma } x^{3/2} }
  e^{- \frac{f^2 x}{ 4  \sigma  } }
\end{eqnarray}

The total probability that it happens with a distance
 $y$ between the two minima
reads
\begin{eqnarray}
D_f(y) = \int_0^{\infty} dx_1 P(E=0,x_1,x_1+y)
= d_f(y)
=  \frac{ 1 }{\sqrt{ 4 \pi \sigma } y^{3/2} }
  e^{- \frac{f^2 y}{ 4  \sigma  } }
\end{eqnarray}
In particular, the second moment reads
\begin{eqnarray}
\int_0^{\infty} dx x^2 {\cal D}_f(x)
=  \frac{ 2 \sigma }{f^3}
\end{eqnarray}
and thus corresponds exactly to the result (\ref{resbiased}) via (\ref{identitylowT}).

The relation (\ref{momentslowT}) for the moments yields
\begin{eqnarray}
 \overline {  < (x-<x>)^{2k} > } && = \frac{T}{k} \int_0^{+\infty} y^{2k} D(y)
 =  \frac{T}{k}  \frac{ \Gamma(2k-\frac{1}{2} )  }{  \sqrt{ 4 \pi \sigma    } }
\left( \frac{ 4 \sigma}{ f^2} \right)^{2k-1/2}
\end{eqnarray}
whereas the relation (\ref{chaosordret}) reads
\begin{eqnarray}
\overline{ (< x >_{T} - < x >_{T=0})^2 } =
T   (2 \ln 2-1) \frac{ 2 \sigma }{f^3}  +O(T^2)
\end{eqnarray}

\subsection{Conclusion}

We have shown on various explicit examples
(the toy model \cite{us_toy}, the pure Brownian potential on a finite sample
in \ref{brownian}, or the biased Brownian potential in \ref{drift})
how the thermal properties equilibrium properties of a particle
in one-dimensional potentials can
be obtained at order $T$ in temperature
from the statistical properties of configurations with
two nearly degenerate ground states.
We will now generalize this approach
to a model with many degrees of freedom.

\section{Low-temperature properties of the 1D random-field Ising model }

\label{secrfim}

\subsection{ Model and notations}

\begin{figure}[b]
\centerline{\includegraphics[height=4cm]{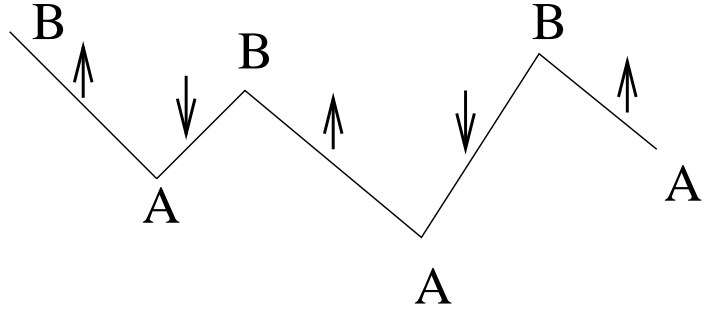}} 
\caption{ Picture of the ground-state : the zig-zag line represents the renormalization
of the random potential $V(x)=2 \sum_{i=1}^x h_i$ seen by the domain walls,
where only extrema separated by $\Delta V> 4 J$ have been kept.
Then all maxima are occupied by a domain wall $B(-\vert+)$, and all
minima are occupied by a domain wall $A(+\vert -)$.
So descending bonds are domains of $+$ spins, whereas ascending bonds are
 domains of $-$ spins  } 
\label{rfimfig}
\end{figure}

The Hamiltonian of the one-dimensional  random-field Ising chain reads
\begin{eqnarray}
{\cal H} = - J \sum_{i=1}^{N-1} S_i S_{i+1} - \sum_{i=1}^{N} h_i S_i
\label{rfim}
\end{eqnarray}
where the $h_i$ are independent random variables with
zero mean value $\overline{h_i}=0$ and variance
\begin{eqnarray}
\overline{h_i^2} = g
\end{eqnarray}
In the following we will moreover assume that the distribution of $h_i$
is continuous to avoid the presence of an extensive number of
degenerate ground states.
The equilibrium of this model can be studied in details
via a Ma-Dasgupta real-space RG approach \cite{us_rfim}
in the universal regime where the Imry-Ma length \cite{imryma},
which represents the typical size of domains
in the ground-state of (\ref{rfim})
\begin{eqnarray}
L_{IM} \equiv \frac{4 J^2}{g} \gg 1
\label{imryma}
\end{eqnarray}
is large. A picture of the ground state is given on Figure (\ref{rfimfig}).
The essential result that will be useful in the following is
that the join distribution of the length $l$
and the energy gain $F= 2 \vert \sum_{j=i}^{i+l} h_j\vert$
of an Imry-Ma domain reads in Laplace transform \cite{us_rfim}
\begin{eqnarray}
\int_0^{+\infty} dl e^{-p l} P_{\Gamma_J}(F,l) =
\theta(F \geq \Gamma_J) U_{\Gamma_J}(p) e^{- (F-\Gamma_J) u_{\Gamma_J}(p) }
\label{fixedp}
\end{eqnarray}
where
\begin{eqnarray}
&& \Gamma_J=4J \nonumber \\
&& U_{\Gamma}(p) =  \frac{  \sqrt{ \frac{p}{2 g} } }
{ \sinh \Gamma \sqrt{ \frac{p}{2 g} } } \nonumber \\
&& u_{\Gamma}(p) =  \sqrt{ \frac{p}{2 g} }
 \coth \Gamma \sqrt{ \frac{p}{2 g} }
\label{notarfim}
\end{eqnarray}
The average length of domain walls is then exactly $L_{IM}$
\begin{eqnarray}
\overline{l} = \int dF \int dl l P_{\Gamma_J}(F,l) = \frac{\Gamma_J^2}{4 g} =L_{IM}
\end{eqnarray}
i.e. the density of domain walls is simply $n_{DW} = 1/L_{IM}$.

\subsection{ Statistics of nearly degenerate excitations  }

The thermal excitations of nearly vanishing energy
are rare events of two kinds : they have been called rare events of type (a)
and of type (c) respectively in the previous study on aging dynamical properties \cite{us_rfim}.
 These two possibilities are the following :

\begin{figure}[b]
\centerline{\includegraphics[height=6cm]{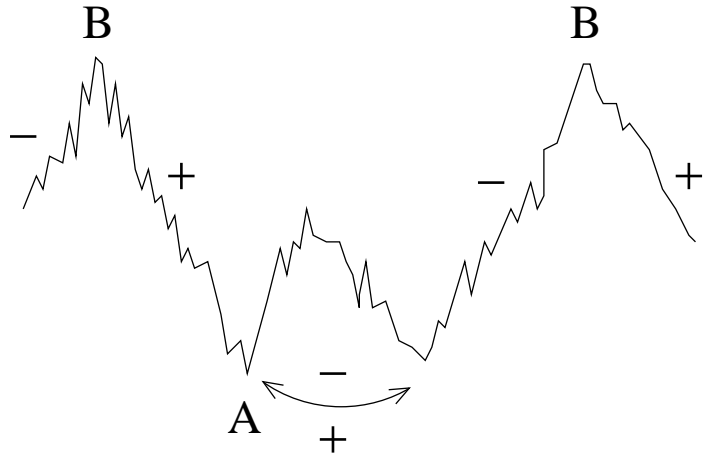}} 
\caption{ Representation of a two-level excitation of type (a) :
a domain wall $A$ of the ground state may have two nearly degenerate optimal positions,
separated by a distance $l$ if $\Delta V = 2 \sum_{i=1}^l h_i \sim 0$.   } 
\label{rarea}
\end{figure}

(i) The excitations of type (a) involve a single domain wall
which has two almost degenerate optimal positions : see Figure \ref{rarea}.
From \cite{us_sinai,us_rfim}, we can obtain the
probability density $\rho_a(E=0,l)$ of these excitations
in the scaling region $l \sim L_{IM} \gg 1$
in terms of (\ref{fixedp})
\begin{eqnarray}
\rho_a^{large}(E=0,l)  = \frac{1}{  L_{IM} }
\int_{\gamma}^{\Gamma} d \Gamma_0 \int_0^{+\infty} dl_1 P_{\Gamma_0}(\Gamma_0,l_1)
  \int_0^{+\infty} dl_2
P_{\Gamma_0}(\Gamma_0,l_2) \delta(l-(l_1+l_2))
\label{rhoaldef}
\end{eqnarray}
where $\gamma$ is a small cut-off.
Indeed, as in the discussion of the Brownian potential,
the probability of degenerate minima diverges at small distance
in the continuum (\ref{dybrownian}), but the moments
that appear in physical observables will be well defined.
Indeed, in Laplace transform, we have using (\ref{notarfim})
\begin{eqnarray}
\int_0^{+\infty} dl  e^{-p l} \ l \ \rho_a^{large}(E=0,l) &&  = \frac{1}{  L_{IM} }
(-\partial_p) \int_{0}^{\Gamma} d \Gamma_0 U_{\Gamma_0}^2(p)
= \frac{1}{  L_{IM} } \partial_p u_{\Gamma}(p)  \\
&& = \frac{1}{  L_{IM} \Gamma_J} \left[ \frac{2}{3} L_{IM} - \frac{8}{45} L_{IM}^2  p +O(p^2)\right]
\label{rhoap}
\end{eqnarray}
These excitations have thus indeed a length of order $l \sim L_{IM}$,
but concerns only a small fraction of order $1/\Gamma_J$
of the domain walls.
After Laplace inversion, the length
distribution is given by an infinite sum of exponentials
\begin{eqnarray} 
\rho_a^{large}(E=0,l) dl = \frac{\pi^2}{ \Gamma_J L_{IM}^2 } 
 \sum_{n=1}^{+\infty} n^2 e^{- n^2 \pi^2 \frac{  l }{ 2 L_{IM}  } }
\label{rhoal}
\end{eqnarray}

Here, since we have a discrete model, the probability of degeneracy
is regularized by the lattice at small distances.
Indeed, nearly degenerate excitations of small length $l \sim 1$
also exist. They are not described by the universal scaling
regime (\ref{rhoal}), but depend on specific properties
of the initial distribution $P(h_i)$.
For instance, the excitations of length $l=1$ correspond to the domain walls
that have a neighbor with a small random field $  h_i  <T \to 0$.
The density of such excitations
is simply proportional to the density $n = 1/L_{IM}$ 
of domain-walls in the ground-state, and to the weight $P(h_i=0)$ of the initial
distribution at the origin
\begin{eqnarray}
\rho_a(E=0,l=1) = \frac{1}{L_{IM} } 2 P(h_i=0)
\label{small1}
\end{eqnarray}
More generally, the statistics of small excitations $l=2,3,..$
 is governed by the probabilities of returns to the origin
for a constrained sum of $l$ of random variables.
For instance, the density of two-spin excitations reads,
assuming the symmetry $P(-h)=P(h)$
\begin{eqnarray}
\rho_a(E=0,l=2) = \frac{1}{L_{IM} } 2 \int_{0}^{+\infty} dh P^2(h)
\label{small2}
\end{eqnarray}

\begin{figure}
\centerline{\includegraphics[height=6cm]{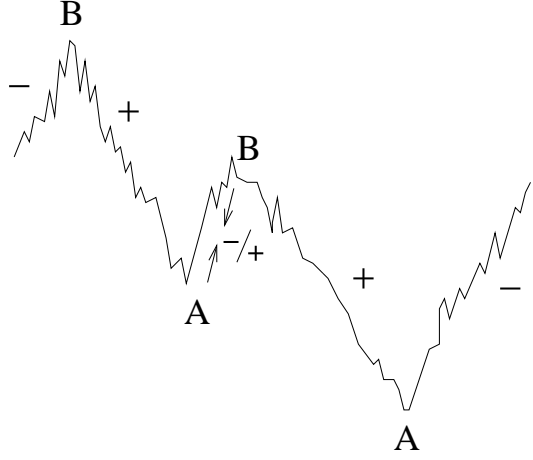}} 
\caption{ Representation of a two-level excitation of type (c) :
a pair (A,B) of neighboring domain walls separated by a distance $l$
may appear or annihilate with almost no energy cost 
if $\Delta V = 2 \sum_{i=1}^l h_i \sim 4 J$.  } 
\label{rarec}
\end{figure}

(ii) The excitations of type (c) involve a pair of domain walls which can appear or annihilate with almost no energy cost : see Figure \ref{rarec}.
These excitation have by definition a large length $l \sim L_{IM} \gg 1$.
From \cite{us_sinai,us_rfim}, their density reads
in terms of (\ref{fixedp})
\begin{eqnarray}
 \rho_c(E=0,l) = \frac{1}{L_{IM} } P_{\Gamma_J} (\Gamma_J,l)
\label{rhocldef}
\end{eqnarray}
i.e. in Laplace transform, we have more explicitly using
(\ref{notarfim})
 \begin{eqnarray}
\int_0^{+\infty} dl  e^{-p l} \rho_c(E=0,l) = \frac{1}{L_{IM} } U_{\Gamma_J} (p)
 = \frac{1}{ \Gamma_J L_{IM} } \left[ 1- \frac{1}{3} L_{IM} p
+ \frac{7}{90} L_{IM}^2 p^2 +O(p^3) \right]
\label{rhocp}
\end{eqnarray}
As in (\ref{rhoap}), these excitations of length $l \sim L_{IM}$
 concern only a small fraction of order $1/\Gamma_J$
of the domain walls. Again, after Laplace inversion, the length
distribution is given by an infinite sum of exponentials
\begin{eqnarray} 
\rho_c(E=0,l)  = \frac{\pi^2  }{ \Gamma_J L_{IM}^2 } 
 \sum_{n=1}^{+\infty} (-1)^{n+1} n^2 e^{- n^2 \pi^2 \frac{  l }{ 2 L_{IM}  } }
\label{rhocl}
\end{eqnarray}

In conclusion, there exists on one hand large excitations of length
$l \sim L_{IM} \gg 1$, whose universal probability densities
are given by the explicit expressions (\ref{rhoal},\ref{rhocl}),
and there are on the other hand small excitations of length $l \sim 1,2,..$, whose statistics
depend on the initial random field distribution,
with the weights (\ref{small1}, \ref{small2}) for $l=1$ and $l=2$ for instance.

\subsection{ Effective Hamiltonian at order $T$ in temperature }

We note $ {\cal C}_0 = \{ S_i^{(0)} \}$ the true ground state,
and we index by $\alpha=1,2,..,l$ the possible strings where the spins
$(S_{i_{\alpha}},...,S_{j_{\alpha}})$ can be coherently returned with
a small amount of energy $E_{\alpha} \sim T \to 0$, as a consequence
of the rare excitations (a) or (c) described above.
These excitations appear with a small probability at low temperature,
i.e. their are dilute, and the
 effective hamiltonian at low-temperature is thus given by
\begin{eqnarray}
e^{-\beta h_{eff} } = e^{-\beta E_0 }\left( \prod_{j \in R} \delta(S_j,S_j^{(0)  }) \right)
 \prod_{\alpha=1}^{\alpha_{max}} \frac{ 1 }{ \left( 1+ e^{- \frac{E_{\alpha}}{T}} \right)}
\left( \prod_{i=i_{\alpha}}^{j_{\alpha}}
\delta(S_i,S_i^{(0)})+ e^{- \frac{E_{\alpha}}{T} }
\prod_{i=i_{\alpha}}^{j_{\alpha}}
\delta(S_i,-S_i^{(0)}) \right)
\label{hrffrfim}
\end{eqnarray}
where $E_0={\cal H}({\cal C}_0)$ is the energy of the ground state configuration
and where $R$ is the set of all the spins $j$
that are outside all the strings $\alpha$.

\subsection{ Specific heat at low temperature }

At low temperature, the effective partition function (\ref{hrffrfim})
is thus given by a product over independent two-level excitations with respect to the ground state
\begin{eqnarray}
Z_{eff}(T) \sim e^{- \beta E_0} \prod_{\alpha} \left( 1+ e^{- \frac{E_{\alpha}}{T}} \right)
\end{eqnarray}
 The free-energy per site reads in the thermodynamic limit thus reads
\begin{eqnarray}
f(T) = -T \lim_{N \to \infty}
\frac{1}{N} \overline{ \ln Z_{eff} } = e_0
-T \int_0^{+\infty} dE \rho(E) \ln \left( 1+ e^{- \frac{E}{T}} \right)
\end{eqnarray}
where the ground state energy $e_0= -J -g/(2 J) $ was already discussed in
\cite{us_rfim}, and where $\rho(E)$ represents the probability density
of excitations of all types and of all sizes
\begin{eqnarray}
\rho(E)= \sum_l \left[ \rho_a(E,l) + \rho_c(E,l) \right]
\end{eqnarray}
 The linear term in temperature of the
specific heat is directly related to the density
of excitations at zero-energy
\begin{eqnarray}
C(T)= T \frac{\pi^2}{6} \rho(E=0) +O(T^2)
\label{ctrho}
\end{eqnarray}

\subsection{ Edwards-Anderson order parameter}

We now consider the generating function of the thermal fluctuations
of a given spin $S_k$
\begin{eqnarray}
Z_{eff}(\lambda) \equiv \overline{ \ln < e^{-  \lambda_K S_k} > }
\end{eqnarray}
\begin{eqnarray}
\overline{ \ln Z_{eff} ( \lambda_k  ) } =
 << \sum_{\alpha=1}^{\alpha_{max}}
\theta(i_{\alpha} \leq k \leq j_{\alpha} ) \ln \frac{ \left( 1 + e^{- \frac{E_{\alpha}}{T} }
e^{2  \lambda_k S_k^{(0)} } \right) }{ \left( 1+ e^{- \frac{E_{\alpha}}{T}} \right)} >>
\end{eqnarray}
where $<<..>>$ denotes the average with respect to the properties
of the nearly-degenerate strings $\alpha$. Using now $S_k^{(0)}=\pm 1$
and the results of
(\ref{integralenergy}) for the average over the energies $E_{\alpha}$,
we get
\begin{eqnarray}
\overline{ \ln < e^{-  \lambda_k S_k} > } =
T \lambda_k^2 \int_0^{+\infty} dl \ l \rho(E=0,l) +O(T^2)
\label{genespink}
\end{eqnarray}
The thermal fluctuations of a spin are thus directly
proportional to the probability
$\int_0^{+\infty} dl \ l \rho(E=0,l)$
that the spin belongs to a nearly degenerate excitation.
In particular, since $S_k^2=1$, the Edwards-Anderson order parameter reads
\begin{eqnarray}
q_{EA} \equiv \overline{ <S_k>^2 } =1 -   2 T
 \int_0^{+\infty} dl \ l \rho(E=0,l) +O(T^2)
\label{qea}
\end{eqnarray}
The vanishing of higher cumulants in (\ref{genespink}) yields
further identities, the first ones being for instance
\begin{eqnarray}
&& \overline{ <S_k>^4 } = \frac{4 \overline{ <S_k>^2 } -1 }{3}  +O(T^2)
= 1-   \frac{8}{3} T
 \int_0^{+\infty} dl \ l \rho(E=0,l) +O(T^2) \\
&& \overline{ <S_k>^6 } = \frac{ 30 \overline{ <S_k>^4 } - 17 \overline{ <S_k>^2 } +2  }{ 15} +O(T^2) = 1 -   \frac{46}{15}  T
 \int_0^{+\infty} dl \ l \rho(E=0,l) +O(T^2)
\end{eqnarray}

\subsection{ Susceptibility }

We now consider the global magnetization
\begin{eqnarray}
m= \frac{1}{N} \sum_{i=1}^N  S_i
 \end{eqnarray}
and the generating function of its thermal fluctuations
\begin{eqnarray}
Z_{eff}(\lambda) \equiv \overline{ \ln < e^{- \lambda m} > }
\label{genespin}
\end{eqnarray}
The effective Hamiltonian at order $T$ in temperature (\ref{hrffrfim})
yields the following partition function with source $\lambda$
\begin{eqnarray}
Z_{eff}( \lambda  ) = \sum_{S_i=\pm 1} e^{-\beta h_{eff}-  \lambda m  } = e^{- \frac{\lambda}{N} \sum_{i=1}^N  S_i^{(0)}  }
 \prod_{\alpha=1}^{\alpha_{max}}  \frac{
\left( 1 + e^{- \frac{E_{\alpha}}{T} }
e^{2 \frac{\lambda}{N} \epsilon_{\alpha} l_{\alpha}  } \right) }
{ \left( 1+ e^{- \frac{E_{\alpha}}{T}} \right)}
\label{argrfim}
\end{eqnarray}
where $\epsilon_{\alpha}=sgn(S_i^{(0),\alpha})=\pm 1$ with probabilities $(\pm 1)$.
We thus obtain
\begin{eqnarray}
\overline{ \ln Z_{eff} ( \lambda  ) } =
 << \sum_{\alpha=1}^l \ln \frac{ \left( 1 + e^{- \frac{E_{\alpha}}{T} }
e^{2 \frac{\lambda}{N} \epsilon_{\alpha} l_{\alpha} } \right) }{ \left( 1+ e^{- \frac{E_{\alpha}}{T}} \right)} >>
\end{eqnarray}
where $<<..>>$ denotes the average with respect to the properties
of the nearly-degenerate strings $\alpha$. Using now the results of
(\ref{integralenergy}) for the average over the energies $E_{\alpha}$,
we finally get
\begin{eqnarray}
\overline{ \ln Z_{eff} ( \lambda  ) } = T  \frac{\lambda^2}{N}
\int_0^{+\infty} dl \ l^2 \rho(E=0,l) +O(T^2)
\end{eqnarray}

In particular, the average susceptibility reads
\begin{eqnarray}
\overline{\chi} \equiv \frac{N}{T} \left( \overline{ <m^2>-<m>^2 } \right)
= 2
\int_0^{+\infty} dl \ l^2 \rho(E=0,l) +O(T)
\label{meanchi}
 \end{eqnarray}
so the excitations appear here with a weight $l^2$.

The susceptibility (\ref{meanchi}) we have just obtained from the analysis
of the thermal fluctuations of the magnetization at
linear order in $T$ may also be recovered from
the analysis of the response to an external field $H$ at zero temperature.

Indeed, suppose we start from the ground state at $H=0$, and we turn on
 a small external field $H>0$. The changes in the ground state will
be localized on the nearly degenerate excitations of small energy $E>0$
and length $l$ such that the new energy in field allows to decrease the energy $E_H=E - 2 H l <0$ : the flip of this excitation yields a change of $m=l$
in the magnetization. As a consequence, the total magnetization per spin
is given by
\begin{eqnarray}
M(H,T=0) = \int_0^{+\infty} dl \int_0^{2 H l} dE \rho(E,l) l
\opsimeq_{H \to 0} 2 H
\int_0^{+\infty} dl \ l^2 \rho(E=0,l)
 \end{eqnarray}
in agreement with (\ref{meanchi}), as it should to recover the Fluctuation-Dissipation theorem.

\subsection{ Discussion : the role of
small/large excitations on various observables }

Since there are two kinds of excitations,
the small ones $l \sim 1$ and the large ones $l \sim L_{IM}$,
and since various observables are governed by the nearly
degenerate excitations with various weights involving their lengths,
it is now interesting to discuss which observables are dominated by
small excitations, and which observables are governed by large excitations.

To simplify the discussion, and to compare with the existing rigorous results
\cite{theoluck}, we now consider that the initial distribution $P(h_i)$ of the random fields
has the following scaling form
\begin{eqnarray}
P(h_i)= \frac{1}{H} {\cal P} \left(\frac{\vert h_i \vert }{H} \right)
\label{examplescaling}
\end{eqnarray}
where ${\cal P}(x)$ is a continuous function, such as
for instance the exponential distribution ${\cal P}(x)=e^{- x  }/2$
considered in \cite{theoluck} among other cases.
The Imry-Ma length then reads
\begin{eqnarray}
L_{IM}= \frac{4 J^2}{g} = \frac{4 J^2}{ H^2 \int_0^{+\infty} dx x^2 {\cal P}(x)}
\sim  (cte) \frac{ J^2 }{ H^2}
 \end{eqnarray}
The density of large scale excitations scales as
 (\ref{rhoap} and \ref{rhocp})
\begin{eqnarray}
\rho^{large}(E=0) = \rho^{large}_a(E=0)+\rho^{large}_c(E=0) \sim
\frac{ 1 }{  L_{IM} \Gamma_J} = \frac{ g }{ 16 J^3}
\sim \frac{ H^2 }{  J^3}
\label{densitylarge}
\end{eqnarray}
On the other hand, the weight of the smallest excitations $l=1$ and $l=2$
reads (\ref{small1},\ref{small2})
\begin{eqnarray}
\rho_a(E=0,l=1) && = \frac{1}{ L_{IM} } \frac{ 2 {\cal P}(0) }{ H}
\\
\rho_a(E=0,l=2) && = \frac{1}{L_{IM} } \frac{ 2 \int_{0}^{+\infty} dx
{\cal P}^2(x) }{H}
\label{scalsmall}
\end{eqnarray}
So even if the initial distribution has a hole at zero field ${\cal P}(0)=0$,
since the prefactor $\int_{0}^{+\infty} dx
{\cal P}^2(x)$ of excitations of length $l=2$ cannot vanish,
the density of small excitations has always for scaling
\begin{eqnarray}
\rho_{a}^{small}(E=0) = \frac{1}{ H L_{IM} } \sim \frac{ H }{  J^2}
\label{densitysmall}
\end{eqnarray}

In the regime we consider, where the Imry-Ma length is very large
$L_{IM} \gg 1$, the density of small excitations (\ref{densitysmall})
is thus much bigger than the density of large excitations (\ref{densitylarge})
\begin{eqnarray}
\rho_{a}^{small}(E=0) \gg \rho^{large}(E=0)
\end{eqnarray}
As a consequence, the specific heat (\ref{ctrho}) is governed by
the density of these small excitations (\ref{densitysmall})
\begin{eqnarray}
C(T)= T \frac{\pi^2}{6} \rho(E=0) +O(T^2) \sim  T b \frac{ H }{  J^2} + ...
\label{cthj2}
\end{eqnarray}
where the numerical coefficient $b$ depend upon the form
of the initial distribution via the prefactors
of the densities of small excitations (\ref{scalsmall}).
The leading behavior (\ref{cthj2}) is indeed in agreement with the
exact results obtained via the Dyson-Schmidt method for
various models of the form (\ref{examplescaling}).

Let us now consider the Edwards-Anderson order parameter (\ref{qea}) :
the density of large excitations weighted by their lengths
has for expression (\ref{rhoap},\ref{rhocp})
\begin{eqnarray}
 \int_0^{+\infty} dl \ l \left[\rho_a^{large}(E=0,l) +\rho_c(E=0,l) \right]
=  \frac{1}{   \Gamma_J} \left( \frac{2}{3} + \frac{1}{3} \right)
= \frac{1}{   \Gamma_J} = \frac{1}{ 4 J }
\label{contrilargebis}
\end{eqnarray}
whereas the
contribution of small excitations (\ref{small1}, \ref{small2})
has the same scaling as the density of small excitations (\ref{densitysmall})
\begin{eqnarray}
\sum_{l=1,2,\cdots} l \rho_a(E=0,l)  = \rho_a(E=0,l=1) +2 \rho_a(E=0,l=2 ) + \cdots \sim \frac{ H }{  J^2}
 \end{eqnarray}
In the regime we consider, where the Imry-Ma length is very large
$L_{IM} \gg 1$, the Edwards-Anderson order parameter (\ref{qea})
is thus dominated by the large excitations contribution (\ref{contrilargebis})
and reads
\begin{eqnarray}
 1-q_{EA} = 2 T
 \int_0^{+\infty} dl \ l \rho(E=0,l) +O(T^2) \sim \frac{T}{ 2 J } + ..
\end{eqnarray}

Let us now consider the susceptibility (\ref{meanchi}) :
the density of large excitations weighted by the square
of their lengths
has for expression (\ref{rhoap},\ref{rhocp})
\begin{eqnarray}
 \int_0^{+\infty} dl \ l^2 \left[\rho_a^{large}(E=0,l) +\rho_c(E=0,l) \right]
=  \frac{L_{IM}}{   \Gamma_J} \left( \frac{8}{45} + \frac{7}{45} \right)
= \frac{L_{IM}}{ 3  \Gamma_J} = \frac{J}{ 3  g } \sim \frac{J}{ H^2 }
\label{contrilarge}
\end{eqnarray}
whereas the
contribution of small excitations (\ref{small1}, \ref{small2})
has the same scaling as the density of small excitations (\ref{densitysmall})
\begin{eqnarray}
\sum_{l=1,2,\cdots} l^2 \rho_a(E=0,l)  =    \rho_a(E=0,l=1) +4 \rho_a(E=0,l=2 ) + \cdots \sim \frac{ H }{  J^2}
 \end{eqnarray}
In the regime we consider, where the Imry-Ma length is very large
$L_{IM} \gg 1$, the disorder averaged susceptibility (\ref{meanchi})
is thus dominated by the large excitations contribution (\ref{contrilarge})
and reads
\begin{eqnarray}
\overline{\chi} (T=0)
\simeq \frac{2 J}{3 g}
\label{meanchilarge}
 \end{eqnarray}
This result is in agreement with the following exact result
\begin{eqnarray}
\chi(T=0)=  \frac{ 2 J }{3 H^2 } \left[ 1+O \left( \frac{H}{J} \right) \right]
 \end{eqnarray}
obtained via the Dyson-Schmidt method \cite{theoluck} for the
case (\ref{examplescaling}) with the symmetric exponential
distribution ${\cal P}(x)=e^{- x  }/2$.

As a final remark, we may mention the case of discrete
distributions for the initial disorder, such as the binary case
for instance $h_i=\pm h$. The physics is completely different,
because the nearly degenerate excitations that exist for continuous
distributions become now exact degeneracies : the entropy thus become
finite at zero temperature \cite{livreluck}, instead of a unique ground state,
but now there is a gap in low-energy excitations \cite{livreluck}.

\section{Conclusion}

In this paper, we have shown explicitly on two kinds of 1D disordered models
how the analysis of the low-temperature properties in terms of two-level
excitations allows to fully recover the exact results obtained independently.
In particular, we have obtained that the thermal fluctuations of a particle
in 1D random potentials are entirely due to the rare samples presenting
two nearly degenerate minima,
whereas the properties of the random field Ising chain are controlled by
the rare regions that can coherently flip with a vanishing energy cost.
We have also explained how the identities on thermal cumulants
that are valid for
systems presenting a statistical tilt symmetry \cite{identities88}
actually hold more generally for various observables
at order $T$ in temperature, as a result of the statistical properties
of two-level excitations.
Finally, for the random-field Ising chain,
our analysis shed light on the influence of small/large
excitations on various observables :
 the specific heat is dominated by small non-universal
excitations that are more numerous,
whereas the
the susceptibility and the Edwards-Anderson order parameter
are dominated by universal large excitations, whose
length is of order Imry-Ma length : these excitations are rarer
than the small ones, but involve a larger number of spins.

In conclusion, since the phenomenology of two-level excitations
 is very often advocated in the field of disordered systems,
we hope that the present detailed study on 1D models
is useful to strengthen these ideas.
Moreover, since the identification of the sample-dependent two-level
excitations and the characterization of their statistical properties
leads to a very clear picture of the low temperature physics,
we hope that other disordered systems will be analyzed 
in the same way in the future.

\section{Acknowledgments}

It is a pleasure to thank G. Biroli, T. Garel, J. Houdayer, J.M. Luck
and O. Parcollet for useful discussions.

\appendix

\section{ Exact results for the Brownian potential on a finite interval }

\label{purepath}

\subsection{Basic path-integral}

To study the equilibrium of a particle in a Brownian finite sample,
 we need the
following path-integral
\begin{eqnarray}
G(u,l \vert u_0) =  \int_{U(0)=u_0}^{U(l)=u}
{\cal D }U(y) e^{-\frac{1}{4 \sigma } \int_0^{L} dy
\left( \frac{d U} {dy} \right)^2 -q \int_0^{L} dy e^{- \beta U(x)}}
\label{pathG}
\end{eqnarray}
whose Laplace transform with respect to $l$ has been given in \cite{us_golosov}
in terms of Bessel functions
\begin{eqnarray}
\hat{G}(u,p|u_0) && = \frac{2}{ \beta  \sigma }
K_{\nu}\left(\frac{2}{\beta}  \sqrt{\frac{q}{\sigma}} e^{- \frac{ \beta u}{2}} \right) I_{\nu}
\left(\frac{2}{\beta}  \sqrt{\frac{q}{\sigma}} e^{- \frac{ \beta u_0}{2}}
\right)
\qquad \text{if} ~~ - \infty < u \leq u_0 <+\infty \\
\hat{G}(u,p|u_0) && = \frac{2}{ \beta  \sigma }
K_{\nu} \left(\frac{2}{\beta}  \sqrt{\frac{q}{\sigma}} e^{- \frac{ \beta u_0}{2}} \right) I_{\nu}
\left(\frac{2}{\beta}  \sqrt{\frac{q}{\sigma}} e^{- \frac{ \beta u}{2}} \right)
\qquad \text{if} ~~ - \infty < u_0 \leq u <+\infty
\label{resG}
\end{eqnarray}
where $
 \nu=  \frac{2}{\beta}  \sqrt{\frac{p}{\sigma}} $

\subsection{Generating function of the thermal cumulants}

The generating function of the distribution of
the partition function with source (\ref{zlambda})
may be computed in terms of the path-integral (\ref{pathG}) as follows
\begin{eqnarray}
\overline{  e^{-q Z_L(\lambda) } }
&& =  \int_{-\infty}^{+\infty} du \int_{U(0)=0}^{U(L)=u}
{\cal D }U(y) e^{-\frac{1}{4 \sigma } \int_0^{L} dy
\left( \frac{d U} {dy} \right)^2 -q \int_0^{L} dy e^{-\lambda x - \beta U(x)}} \\
&& = e^{- \frac{\lambda^2}{4 \sigma \beta^2} L}
 \int_{-\infty}^{+\infty} dv e^{ \frac{\lambda}{2 \beta \sigma} v}
\int_{V(0)=0}^{V(L)=v}
{\cal D }V(y) e^{-\frac{1}{4 \sigma } \int_0^{L} dy
\left( \frac{d V} {dy} \right)^2 -q \int_0^{L} dy e^{ - \beta V(x)}} \\
&& =  e^{- \frac{\lambda^2}{4 \sigma \beta^2} L}
 \int_{-\infty}^{+\infty} dv e^{ \frac{\lambda}{2 \beta \sigma} v}
G(v,L \vert 0)
\end{eqnarray}
so that in Laplace with respect to $L$, we get using (\ref{resG})
\begin{eqnarray}
\int_0^{+\infty} dL e^{- \omega L} \overline{  e^{-q Z_L(\lambda) } }
&& =    \int_{-\infty}^{+\infty} dv e^{ \frac{\lambda}{2 \beta \sigma} v}
{\hat G} (v, \omega+ \frac{\lambda^2}{4 \sigma \beta^2}  \vert 0)  \\
&& = \frac{4}{ \beta^2  \sigma } T_{\rho,a}(s)
\label{genT}
\end{eqnarray}
where
\begin{eqnarray}
T_{\rho,a}(s) \equiv s^a I_{\rho} (s)
 \int_{s}^{+\infty}  \frac{dz}{z^{1+a}} K_{\rho}(z)
 + s^a K_{\rho}(s) \int_{0}^{s}  \frac{dz}{z^{1+a}} I_{\rho}(z)
\label{defT}
\end{eqnarray}
with the notations
\begin{eqnarray}
s && = \frac{2}{\beta}  \sqrt{\frac{q}{\sigma}} \\
\rho &&= \sqrt{ \nu^2 + a^2}
\end{eqnarray}
and where $a$ and $\nu$ have defined in (\ref{nota}).

To obtain the moments
\begin{eqnarray}
\int_0^{+\infty} dL e^{- \omega L} \overline{  e^{-q Z_L(p) } }
= \sum_{n=0}^{+\infty} \frac{(-q)^n}{n!}
\int_0^{+\infty} dL e^{- \omega L} \overline{ [Z_L(p)]^n  }
\end{eqnarray}
we need to expand (\ref{genT})in $q$, i.e. to expand
the function (\ref{defT}) at all orders in $s$.
The properties of Bessel functions (differential equation and wronskian)
leads to the following differential equation for (\ref{defT}
\begin{eqnarray}
T_{\rho,a}''(s) && = \frac{2a-1}{s} T_{\rho,a}'(s)
+ \left[1 + \frac{\rho^2- a^2}{s^2} \right]T_{\rho,a}(s) -\frac{1}{s^2}
\label{propT}
\end{eqnarray}

Inserting the series expansion
\begin{eqnarray}
T_{\rho,a}(s) =   \sum_{m=0}^{+\infty} \alpha_m s^m
\label{expT}
\end{eqnarray}
into the differential equation (\ref{propT}), one finds $\alpha_{2n+1}=0$
$\alpha_0=1/\mu^2$ and the recurrence
\begin{eqnarray}
 \alpha_{2n} = \frac{ \alpha_{2n-2} }{ (2n-a)^2 -\rho^2 }
\end{eqnarray}
leading to
\begin{eqnarray}
 \alpha_{2n}
= \frac{ \Gamma\left(1- \frac{\rho}{2} - \frac{a}{2}\right)
\Gamma\left(1+ \frac{\rho}{2} - \frac{a}{2}\right) }{
\mu^2 4^n \Gamma\left(n+1- \frac{\rho}{2} - \frac{a}{2}\right)
\Gamma\left(n+1+ \frac{\rho}{2} - \frac{a}{2}\right) }
\label{resal}
\end{eqnarray}

So we obtain the Laplace transform of all moments
\begin{eqnarray}
 \int_0^{+\infty} dL e^{- \omega L} \overline{ [Z_L(p)]^n  }
&& = \left( \frac{4}{\beta^2 \sigma }\right)^{n+1} (-1)^{n} n! \alpha_{2n}  \\
&& = \frac{1}{\omega} \left( \frac{1}{\beta^2 \sigma }\right)^{n}
\Gamma(n+1) \frac{ \Gamma\left(1+ \frac{\rho}{2} - \frac{a}{2}\right)
\Gamma\left( \frac{\rho}{2} + \frac{a}{2} -n \right) }
{\Gamma\left(n+1+ \frac{\rho}{2} - \frac{a}{2}\right)
\Gamma\left( \frac{\rho}{2} + \frac{a}{2}  \right)}
\end{eqnarray}
in agreement with \cite{yor}
and analytic continuation in $n \to 0$ yields the result (\ref{reslaplace})
given in the text.

\end{document}